\documentclass[a4paper]{article}

\newcommand{\xmax}{\ensuremath{X_{\rm max}}}
\newcommand{\lsim}{\mathrel{\hbox{\rlap{\lower.55ex \hbox{$\sim$}} \kern-.3em \raise.4ex \hbox{$<$}}}}
\newcommand{\gsim}{\mathrel{\hbox{\rlap{\lower.55ex \hbox{$\sim$}} \kern-.3em \raise.4ex \hbox{$>$}}}}

\usepackage{icrc2013}

\title{The modeling of the nuclear
composition measurement
performance of the Non-Imaging
CHErenkov Array (NICHE)}

\shorttitle{NICHE Cosmic Ray Composition Modeling}

\authors{
John Krizmanic$^{1}$,
Douglas Bergman$^{2}$,
\& Pierre Sokolsky$^2$
}

\afiliations{
$^1$ Universities Space Research Association, Columbia, Maryland 21044 USA \\
$^2$ Dept. of Physics \& Astronomy, University of Utah, Salt Lake City, Utah 84112 USA \\
}

\email{jkrizmanic@usra.edu}

\abstract{In its initial deployment, the Non-Imaging CHErenkov Array (NICHE) \cite{NICHEcr2012,NICHEicrc} will measure the flux and nuclear composition of cosmic rays from below $10^{16}$ eV to $10^{18}$ eV by using measurements of the amplitude and time-spread of the air-shower Cherenkov signal to achieve a robust event-by-event measurement of \xmax\ and energy. NICHE will have sufficient area and angular acceptance to have significant overlap with TA/TALE, within which NICHE is located, to allow for energy cross-calibration. In order to quantify NICHE's ability to measure the cosmic ray nuclear composition, 4-component composition models were constructed based upon a poly-gonato model of J. H\"{o}randel \cite{polygonato} using simulated
\xmax\  distributions of the composite composition as a function of energy. These composition distributions were then unfolded into individual components via an analysis technique that included NICHE's simulated \xmax\  and energy resolution performance as a function of energy as well as the effects of finite event statistics. Details of the construction of the 4-component composition models and NICHE's ability to determine the individual components as a function of energy are presented.}

\keywords{Cosmic rays, mass composition, modeling}

\begin{document}
\maketitle

%Begin a section.
\section{Motivation}

The NICHE experiment is designed to measure the cosmic ray flux and nuclear composition from $\gsim 10^{15.8}$ eV to $10^{18}$ eV in its initial deployment.  NICHE will employ an array of 69 easily-deployable, non-imaging Cherenkov detectors with 200-m separation in its current baseline design. CORSIKA simulation studies have demonstrated that this design will have an instantaneous aperture of 3.3 km$^2$ sr for $ E \gsim 10^{15.8}$ eV.  This leads to an expected event rate of $\gsim 1500$ events per year above $10^{17}$ eV and thus will have significant overlap with the measurements of the TALE fluorescence detector, leading to cross-calibration between the two measurement techniques.

NICHE is designed to have remarkable \xmax\ and energy resolution: the baseline design has an  \xmax\ resolution is less than 40 g/cm$^2$ at $10^{16}$ eV and decreases to better than 20 g/cm$^2$ above $10^{16.5}$ eV. The energy resolution is better than 15\% at $10^{16}$ eV and improves with increasing energy.  While this performance is exceptional for a sparsely-filled array measuring the properties of air showers, a fundamental question is how this translates into how well an evolving cosmic ray composition can be measured.

\vspace{-4.mm}
\section{CR Composition Modeling}

In order to determine an experiment's CR composition measurement capability, physically-motivated models are needed to quantify how the galactic cosmic ray composition evolves, and eventually is overtaken by an extragalactic component.  A well-parameterized model is the poly-gonato model detailed by J. H\"{o}randel \cite{polygonato}. The model provides energy-dependent spectra for individual elements, from $z=1$ to $z=92$, and is further described by several different assumptions on the modeling of the spectral cutoff above the CR knee, e.g. rigidity or mass dependent. For the models developed in this paper to quantify NICHE performance, the poly-gonato model with a rigidity-based common spectral index for the modeling of the rigidity cutoff is employed, e.g. via Equation 2 and values in Table 1 and Table 6 in \cite{polygonato}.  However, the methodology described in this paper employed to compress the 92 individual element spectra to a 4- and 5-component model could easily be adapted to the other poly-gonato models described in \cite{polygonato} based upon different assumptions of the form of the galactic CR cutoff.

The motivation for compressing a model of individual elemental spectra to ones with 4- or 5-components is \mbox{twofold}.  First, given the resolution of \xmax\ measurements in air showers, the fitting of a measured composition  with a large number of components complicates the interpretation of the results via adding too many free parameters.  This issue is amplified by the large range of the composition fractions on a per element basis in the cosmic radiation. Another motivation is a more practical one: one will simulate air showers in a composition dependent fraction. Since there is a large computational overhead in simulating air showers, if the number of nuclear components is minimized, without loosing sensitivity to the underlying astrophysics, then the air shower simulation problem becomes more tractable.

{\bf 5-Component Composition Model:}
The 5-component composition model assumes a proton, helium, CNO, silicon, and iron components, each constructed using the individual elemental spectra described by Figure 11a in \cite{polygonato}, based on the form $\Phi(E)=\Phi^0 E^{s}$ where s is spectral index.    The proton and helium models are that for $z=1$ and $z=2$ respectively. The CNO ($z=7$) model is based upon summing the $\Phi_Z^0$ terms (Table 1 in \cite{polygonato}) from $3 \le z \le 9$, i.e. $\Phi^0_{CNO}=\sum_{z=3}^{9} \Phi_Z^0$, and uses a single spectral index of -2.676. The silicon ($z=14$) model is given by $\Phi^0_{Si}=\sum_{z=10}^{24} \Phi_Z^0$ and uses a single spectral index of -2.65. The iron model is given by $\Phi^0_{Fe}=1.1 \sum_{z=25}^{27} \Phi_Z^0$  and assuming a single spectral index of -2.59. While \cite{polygonato} also provides a model for the ultra-heavy ($28 \le z \le 92$) component, this is ignored in the models presented in this paper. This ultra-heavy component adds less than 5\% to the total spectrum below $10^{16}$ eV while growing to $\sim$25\% at $10^{17}$ eV and is considered a small contribution to the $z \le 27$ components.

\begin{figure}[t]
  \centering
  \includegraphics[width=0.45\textwidth]{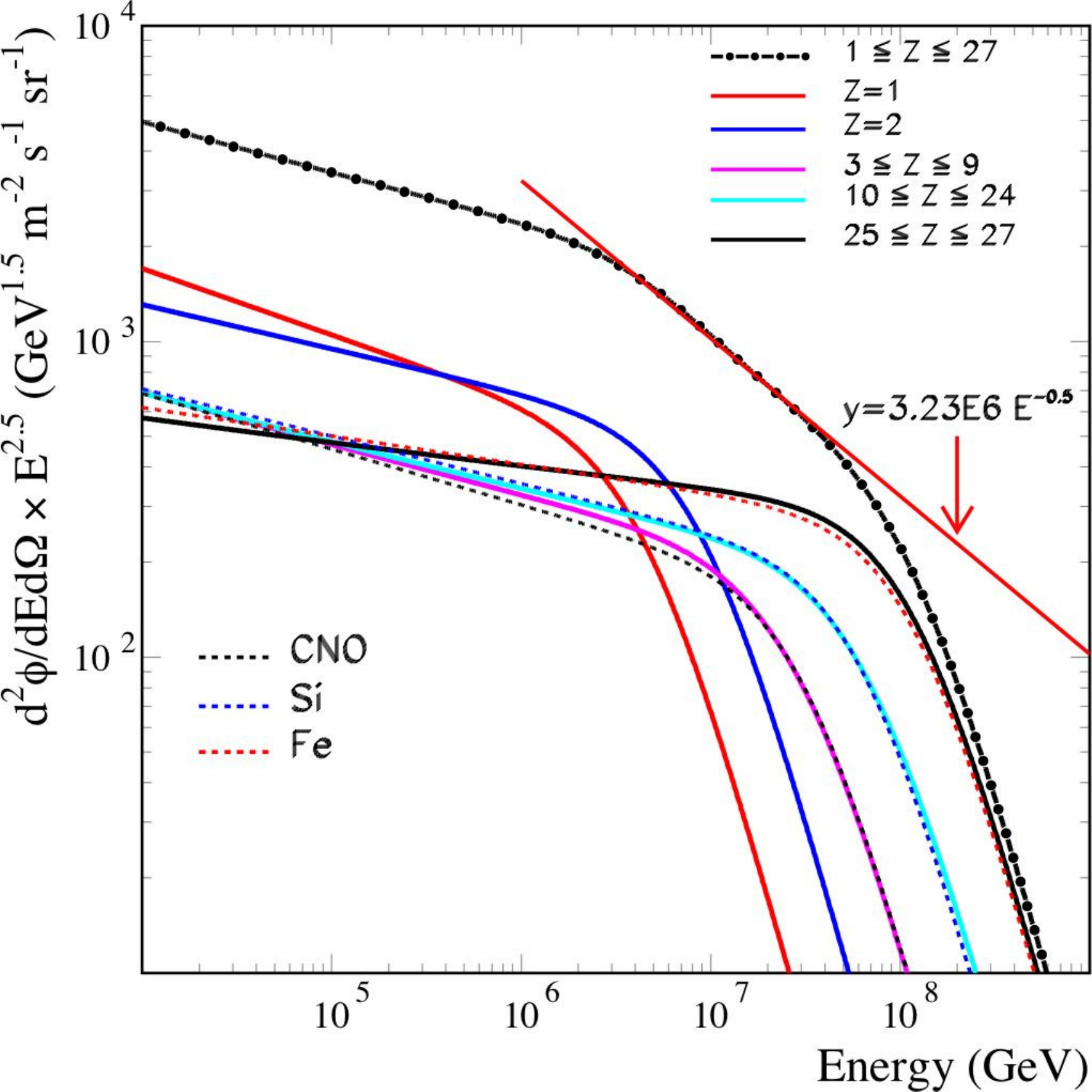}
\vspace{-1mm} 
  \caption{The spectral evolution of the 5-component model (dashed) compared to that predicted by the element-dependent polygonato model.}
  \label{spectevo5_fig}
\vspace{-3mm} 
\end{figure}

\begin{figure}[t]
  \centering
  \includegraphics[width=0.45\textwidth]{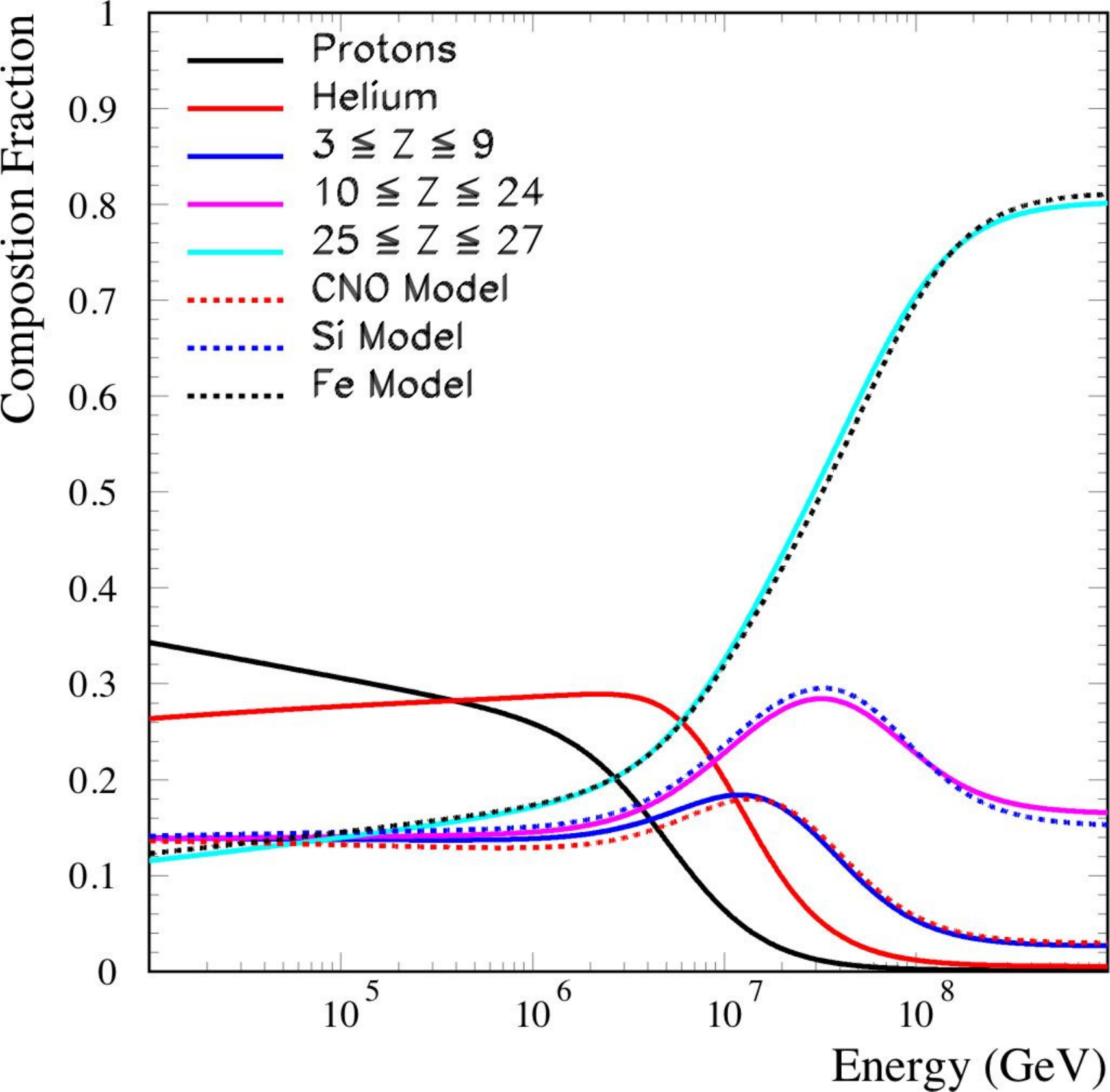}
\vspace{-1mm} 
  \caption{The composition evolution of the 5-component model (dashed) compared to that predicted by the element-dependent polygonato model.}
  \label{compevo5_fig}
\vspace{-3mm} 
\end{figure}

Figure \ref{spectevo5_fig} shows the energy-weighted spectra of the 5 components as compared to that obtained by summing the individual elemental spectra in the appropriate z ranges, except for the proton and helium components that are identical to that for $z=1$ and $z=2$. The composition fraction as a function of energy is shown in Figure \ref{compevo5_fig}.  Note the exceptional agreement between the CNO, Si, and Fe models as compared to there counterparts built from summing the relevant elemental spectra.  The all particle spectra is also shown in Figure \ref{spectevo5_fig} along with an energy-weighted power law spectrum given by $E^{2.5} \phi(E) = 3.23 \times 10^{6} E^{-0.5}$ 
in units of particles $\cdot$ GeV$^{1.5}$ $\cdot$ m$^{-2}$ $\cdot$ s$^{-1}$ $\cdot$ sr$^{-1}$,
which is consisted with the all particle spectrum above $10^{16}$ eV as reported in \cite{polygonato}.

{\bf 4-Component Composition Model:}
The 4-component composition model assumes a proton, helium, CNO, and iron components, each constructed using the a similar procedure used to construct the 5-component model.   The proton, helium and CNO ($z=7$) models are identical to that used in the 5-component model. The new iron model is given by $\Phi^0_{Fe}=0.65 \sum_{z=10}^{27} \Phi_Z^0$  and assuming a single spectral index of -2.59. As with the 5-component model the ultra-heavy ($28 \le z \le 92$) component is ignored. 

\begin{figure}[t]
  \centering
  \includegraphics[width=0.45\textwidth]{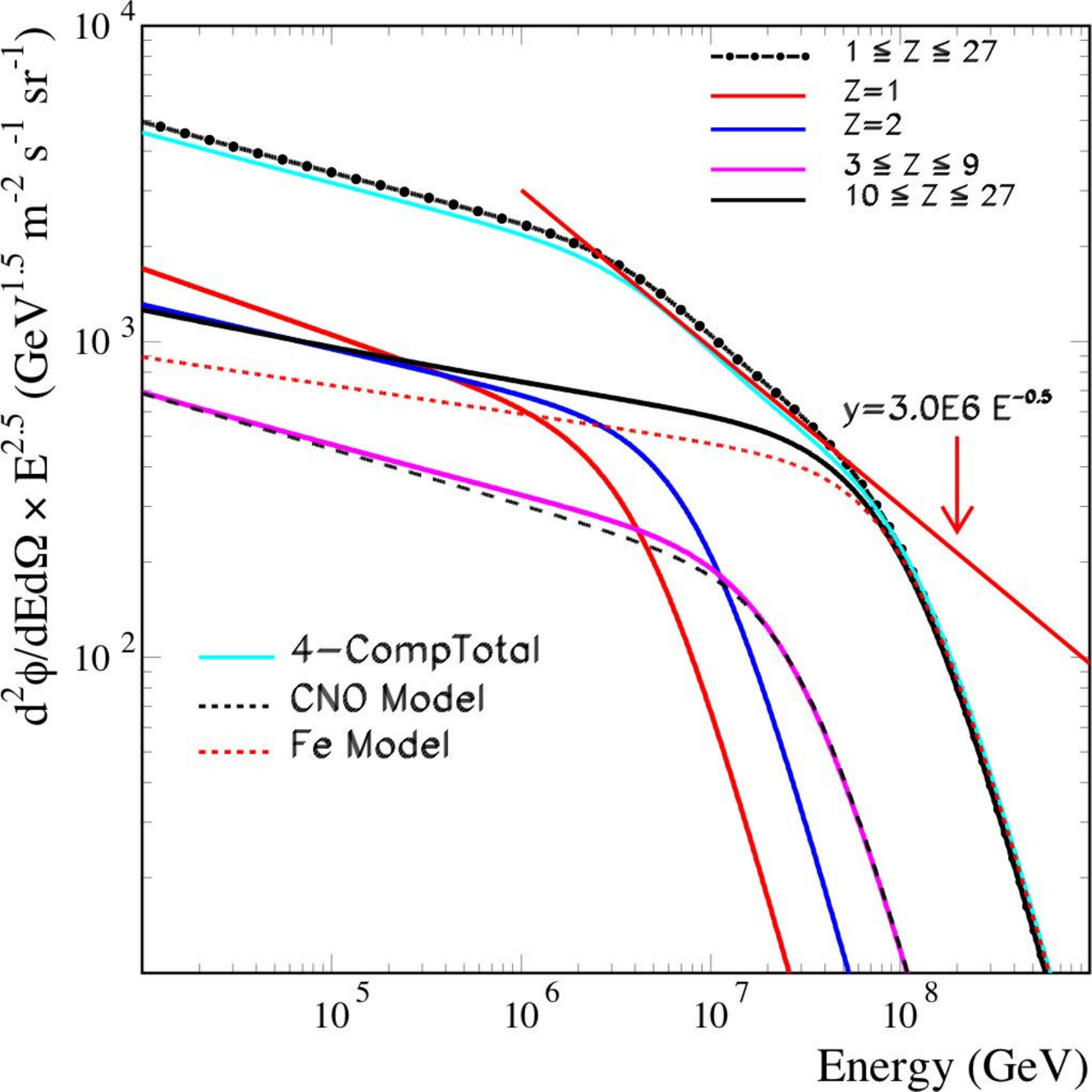}
\vspace{-1mm} 
  \caption{The spectral evolution of the 4-component model (dashed) compared to that predicted by the element-dependent polygonato model.}
  \label{spectevo4_fig}
\vspace{-3mm}  
\end{figure}

\begin{figure}[t]
  \centering
  \includegraphics[width=0.45\textwidth]{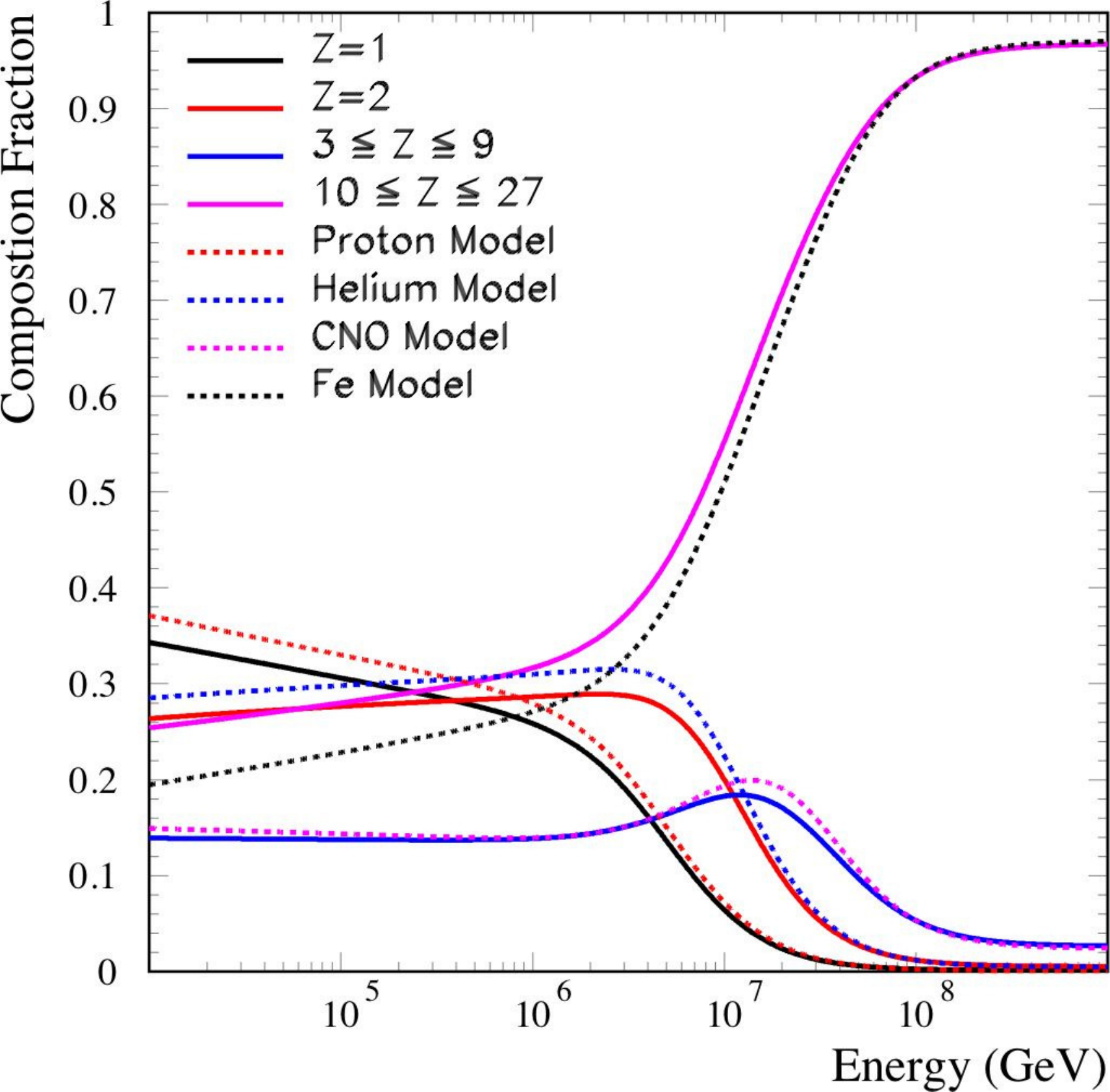}
\vspace{-1mm}   
 \caption{The composition evolution of the 4-component model (dashed) compared to that predicted by the element-dependent polygonato model.}
  \label{compevo4_fig}
\vspace{-1mm}  
\end{figure}

Figure \ref{spectevo4_fig} shows the spectra of the 4 components as compared to summing the individual elemental spectra in the appropriate z ranges, with the proton and helium components identical to the reference spectra. 
The all particle spectra is also shown in Figure \ref{spectevo4_fig} along with an energy-weighted power law spectrum given by $E^{2.5} \phi(E) = 3.0 \times 10^{6} E^{-0.5}$ 
in units of particles $\cdot$ GeV$^{1.5}$ $\cdot$ m$^{-2}$ $\cdot$ s$^{-1}$ $\cdot$ sr$^{-1}$,
which is slightly lower than the all particle spectrum above $10^{16}$ eV reported in \cite{polygonato}. This energy-weighted power law better matches the sum of all 4 components, which is less than that base on summing the individual elemental spectra for $z \le 27$.
The composition fraction as a function of energy is shown in Figure \ref{compevo4_fig}.  Each of the 4 components is slightly different than that based upon using individual elemental components, due to using only 4 components, and modeling the Fe component over a more extended z range.  The fact that the 4 component model slightly diverges the results predicted by using the individual elemental spectra of the reference poly-gonato model highlights the impact of compressing the elemental spectra into 4 components.
However, the 4 component composition model provides a physically meaningful prediction of the composition evolution to be used as a basis for spectra generation and testing how well NICHE can reconstruct the input spectra.

\vspace{-4.mm}
\section{NICHE Response Modeling}

One method for deconvolving distributions formed from multiple components is to use high-statistics individual distributions as templates to be used in a multi-component fitting procedure.  This methodology allows for the effects of instrument response, specifically \xmax\ resolution, to be incorporated both when the all-particle, composite \xmax\ distribution is formed and for the individual component templates. Practically, this procedure can be performed using histograms with binning that slightly oversamples the inherent \xmax\ resolution.  The HBOOK routine HMCMLL \cite{HBOOK} is a specifically designed algorithm to use a log-likelihood fitting procedure to determine the composition of experimental data based upon Monte Carlo simulations of multiple sources that form the composite distribution.

A 1-dimension air shower simulation \cite{Mikulski99} was employed to generate high-statistics samples of \xmax\ distributions for proton, helium, CNO, and iron primary air showers. Composite distributions were then binned in energy with the energy-dependent composition fractions were taken from the 4-component poly-gonato model. The statistics assuming two years of NICHE operation and 10\% duty cycle. Each component was formed by randomly selecting from a library of \xmax\ distributions and then modified to include the effects of \xmax\ and energy resolution as determined by the NICHE performance studies based on CORSIKA simulation studies \cite{NICHEicrc}.

\begin{figure}[t]
  \centering
  \includegraphics[width=0.46\textwidth]{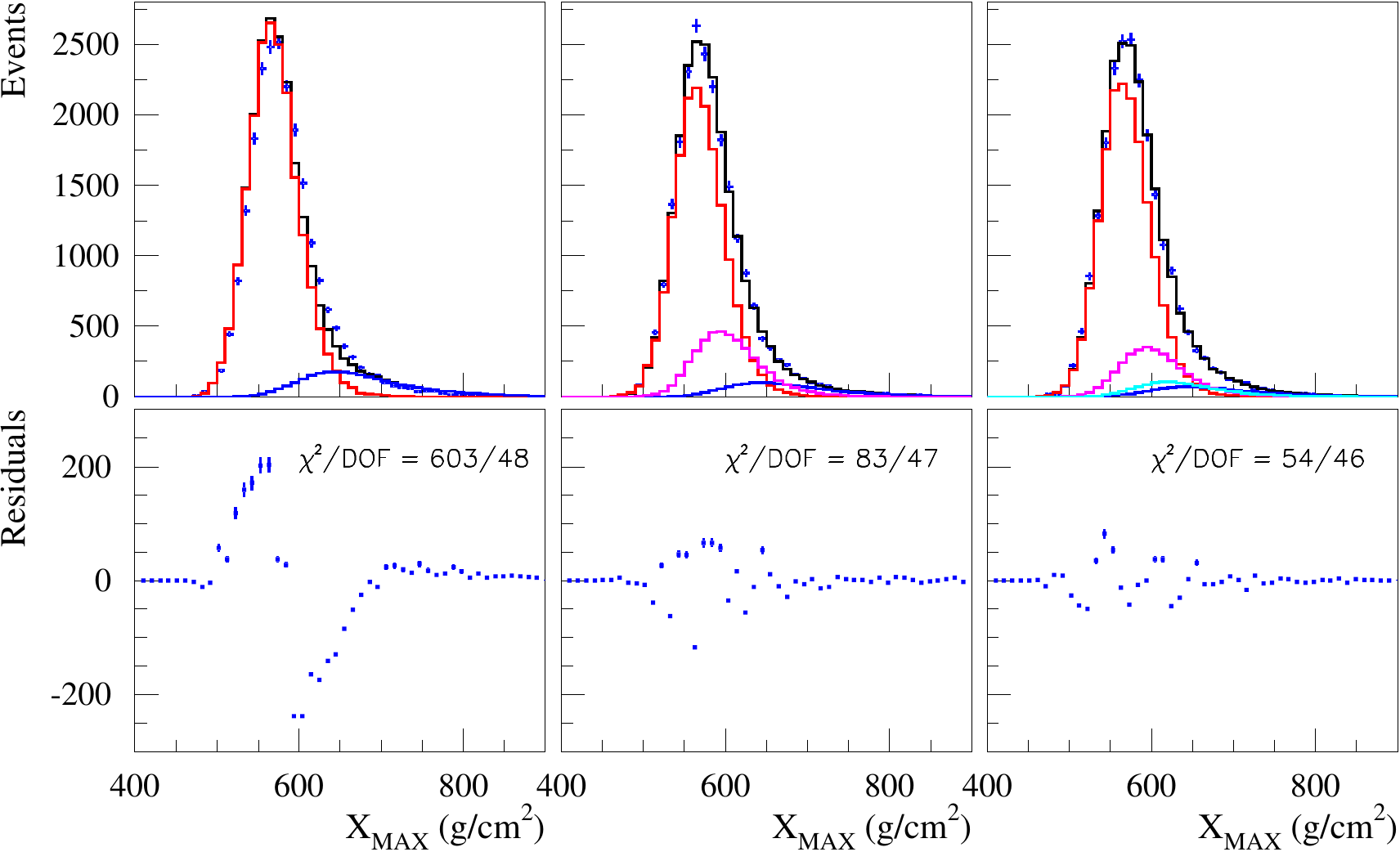}
\vspace{-1mm} 
  \caption{Fits to a simulated HECR composition at $10^{16.5}$
    eV.  22,400 events were selected from a distribution composed of
    5\% protons, 5\% He, 15\% Ni and 75\% Fe.  The upper plots show the results of a 2-, 3-, and 4-component fit while the lower plots show the residuals and reduced $\chi^2$.}
  \label{moneyplot}
\vspace{-3mm} 
 \end{figure}

The HBOOK routine HMCMLL was then used to unfold the composite composition using individual component templates based upon the entire statistics of each component \xmax\ library, but diluted by the predicted \xmax\ resolution. This procedure effectively reconstructs the composition fraction, assuming a 4-component model, of the composite \xmax\ distribution.  Figure \ref{moneyplot} presents the results of this procedure at a specific energy ($10^{16.5}$ eV). Based a two year full exposure estimate for NICHE, 22,400 events were selected from a distribution composed of 5\% protons, 5\% He, 15\% CNO and 75\% Fe.  An \xmax\ resolution of 20 g/cm$^2$ was used to smear the composite \xmax\ distribution on a bin-by-bin basis via a Gaussian with $\sigma=20$ g/cm$^2$. The unfolding procedure then fit the composite distribution used a 2-model, 3-model, and 4-model fitting procedure.

In the left panel of  Figure \ref{moneyplot}, only protons and Fe templates were used for the component fit leading to the large $\chi^2$ value for the fit and the large residuals.  The center panel shows the result of a three-component fit using \mbox{protons}, CNO, and Fe.  The residuals (lower plots) are much
improved, but the $\chi^2$ value is still relatively large.  The right panel shows the 4-component fit and demonstrates a significant improvement in both the residuals and the $\chi^2$. The 4-component fit reconstructs the input composition to within 10\% of the input fractions. The comparison of the results presented in Figure \ref{moneyplot} provides a demonstration of the robustness of the procedure used to unfold the composite \xmax\ distribution into individual components.  That is, a 4-component unfolding accurately recovers the initial composition fractions, while unfolding models with different number of components lead to significantly worse results.

\vspace{-4.mm}
\section{NICHE CR Composition Determination}

Using the 4-component composition model as a basis, \mbox{two} different variations were used to assess the ability of NICHE to distinguish between them.  The first assumes only the predictions for the galactic CR composition while the second adds in a proton component above $10^{16}$ eV to mimic an extragalactic (EG) component.  The EG proton component was taken to be that needed to yield an all-particle flux of
$\phi(E) = 3.0 \times 10^{24} E^{-3}$ particles $\cdot$ eV$^{-1}$ $\cdot$ m$^{-2}$ $\cdot$ s$^{-1}$ $\cdot$ sr$^{-1}$.  For each composition model, composite \xmax\ distributions were formed based upon the energy-dependent component fractions in bins of a third of a decade of energy from $10^{15.8}$ to $10^{17.8}$ eV where NICHE's baseline aperture is constant (see Figure 3 in \cite{NICHEicrc}). The energy-dependent \xmax\ resolution was included for each energy bin, based upon a parameterization of the \xmax\ and energy resolution response determined for well-reconstructed events, currently constrained to events $\le 30^\circ$ zenith angle. The net effect is to reduce the trigger aperture by a factor of 2 to yield a baseline reconstructed aperture. Algorithm development is underway to expand the zenith angle acceptance. The statistics for each energy bin assumes two years of NICHE operation, using the baseline reconstructed aperture and assuming a 10\% duty cycle.  For the energy bin from $10^{17}$ to $10^{17.5}$ eV, this leads to all-particle statistics of approximately 700 events in the galactic-only model and 1200 events in the galactic+EG model.  Above $10^{17.5}$ eV, the two-year baseline NICHE \mbox{statistics} for the galactic+EG model are approximately 170 events, and the modeling of this case was performed for the $10^{17.5}$ to $10^{17.8}$ eV energy bin.  The anticipated statistics in this energy bin for the galactic-only model are too meager to yield a meaningful result. Since NICHE will be operational with TA/TALE, TALE's capabilities to measure composition above $10^{17}$ eV will surpass NICHE at higher energies.  The important point is to have sufficient overlap between NICHE and TALE for cross-calibration.

The procedure is now to generate model- and energy-dependent composite spectra and to unfold these to assess NICHE's capability to reconstruct the thrown composition.  A Monte Carlo technique used the models to generate individual composition fractions in third of a decade energy bins, using \xmax\ distributions from high-statistics \mbox{libraries} defined every tenth of a decade in energy.  These were combined to form the third of a decade results via an  $E^{-3}$ energy-weighting to account for the inherent \xmax\ variation of each energy bin.  The energy-dependent \xmax\ and energy resolutions \cite{NICHEicrc} were included while the anticipated statistics (N) of the composite distributions were Gaussian fluctuated with $\sigma=\sqrt{N}$. The composite distributions were then unfolded assuming a 4-component (p, He, CNO, Fe) model and the component fractions determined.  This was repeated 100 times, randomly choosing each \xmax\ for each event while also independently fluctuating the measurement statistics, yielding distributions of the unfolded fractions defined by a mean and an rms deviation.

\begin{figure}[t]
  \centering
  \includegraphics[width=0.46\textwidth]{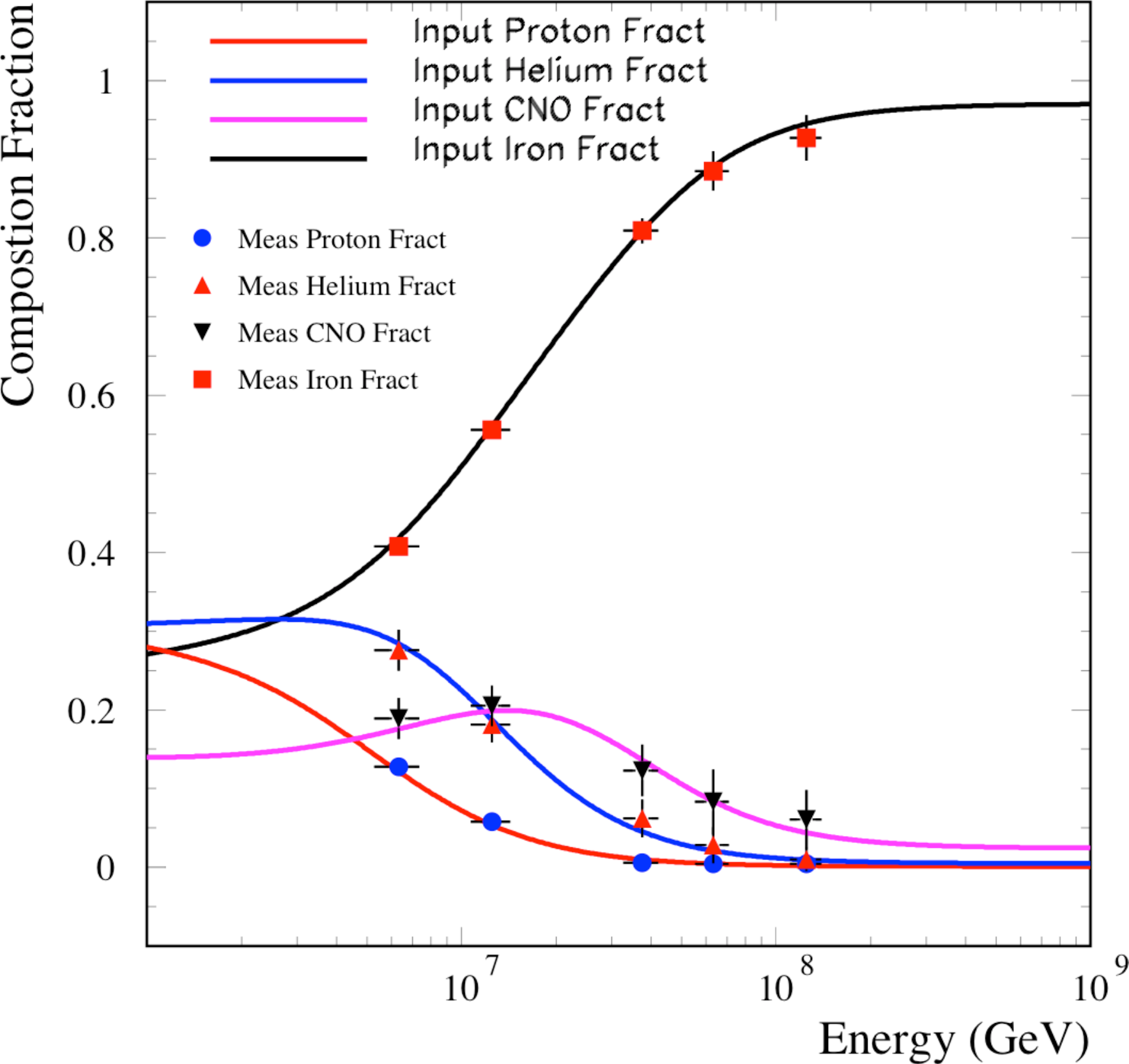}
\vspace{-1mm} 
  \caption{NICHE CR composition reconstruction fraction for CR galactic component only model. The curves represent the generated fractions while the points represent the NICHE measurements based on 2-year statistics.}
  \label{galacticfrac_fig}
\vspace{-3mm}  
\end{figure}

\begin{figure}[t]
  \centering
  \includegraphics[width=0.46\textwidth]{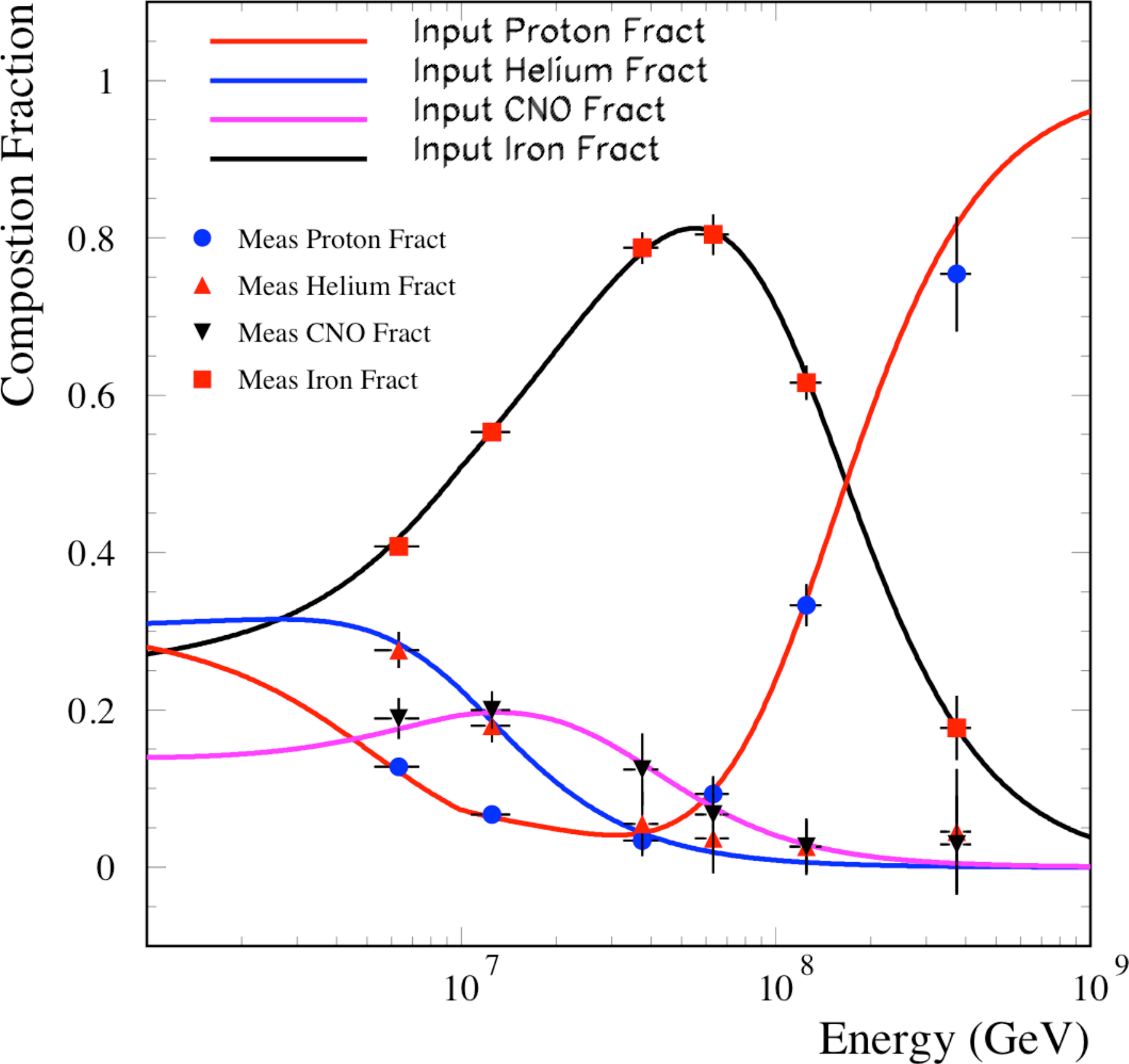}
\vspace{-1mm} 
  \caption{NICHE CR composition reconstruction fraction for CR galactic component \& extragalactic proton model. The curves represent the generated fractions while the points represent the NICHE measurements based on 2-year statistics.}
  \label{extragalactic_fig}
\vspace{-3mm}  
\end{figure}

\begin{figure}
  \centering
  \includegraphics[width=0.47\textwidth]{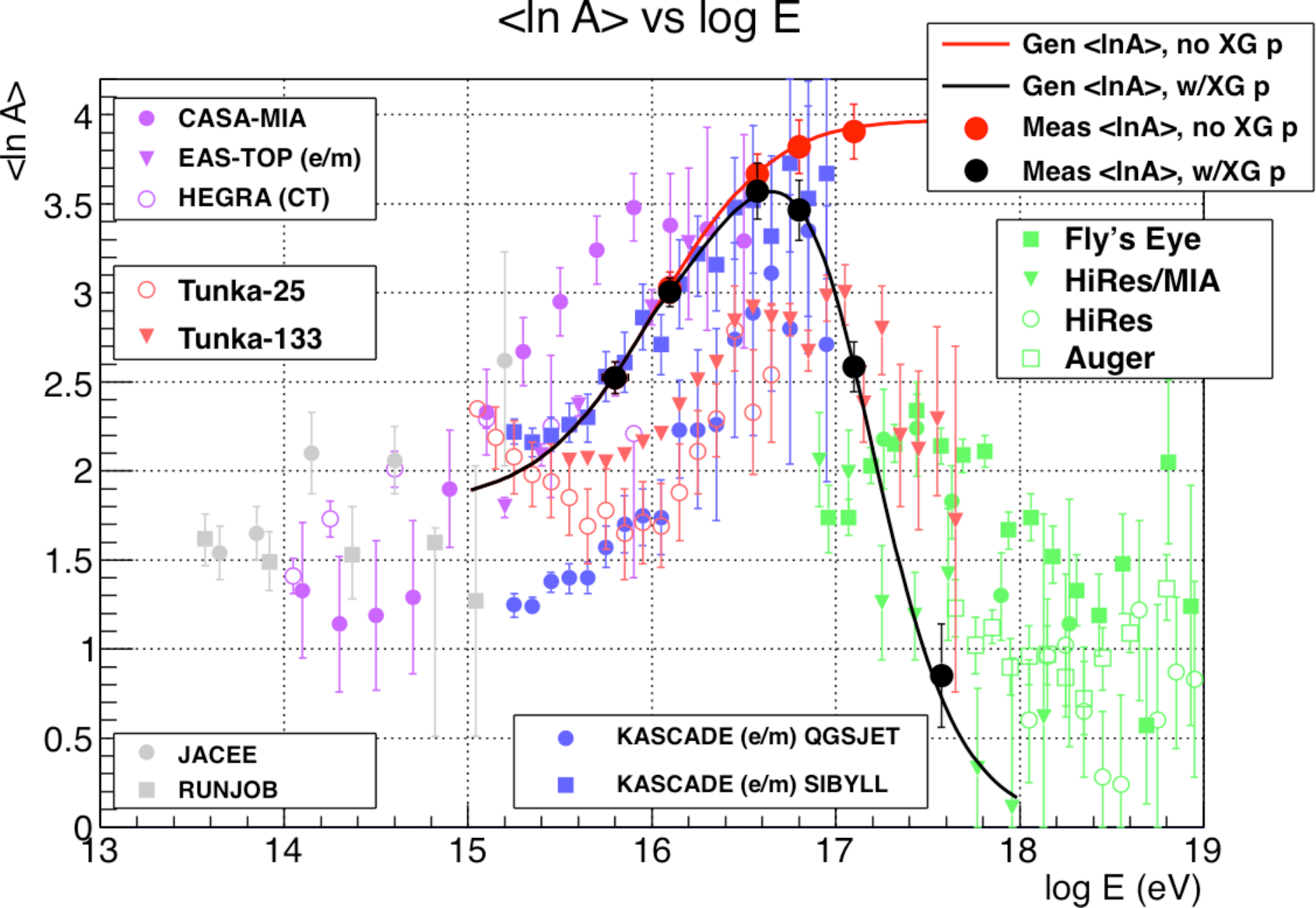}
\vspace{-1mm}  
 \caption{Comparison of NICHE nuclear composition performance compared to to current data (adapted from \cite{Blumer2009}).}
  \label{Ncomp_fig}
\vspace{-3mm} 
 \end{figure}

The results in regards to reconstructing galactic-only and galactic+EG compositions are shown in Figures \ref{galacticfrac_fig} and \ref{extragalactic_fig}.  For component fractions $\gsim 5\%$, NICHE reconstucts the generated composition to $\lsim$ 15\%. The results demonstrate NICHE's ability to accurately measure composition from $10^{15.8}$ to $10^{17.8}$ eV based upon NICHE's baseline response.  Figure \ref{Ncomp_fig} presents the results in terms of $<\ln{A}>$ as compared to a recent data compilation.

\vspace{-4.mm}
\section{Conclusions}

4- and  5-component CR nuclear composition models are presented based upon a poly-gonato model and have been used to quantify ability of the NICHE array to distinguish between two disparate models. These simulation studies indicate that the \xmax\  and energy resolution performance of NICHE will allow for the determination of at least a 4-component model of CR nuclear composition from $10^{15.8}$ up to $10^{17.8}$ eV, i.e. in the region where the galactic CR spectrum is hypothesized to be overtaken by an extragalactic proton-like component.   These studies indicate that NICHE will have sufficient event statistics above $10^{17}$ eV to allow for a cross-calibration between NICHE's Cherenkov air shower measurements to that obtained with TALE's fluorescence measurements, assuming 2 year operation.

\end{document}